\documentclass[preprint,number,sort&compress]{elsarticle}

\usepackage{mathptmx}       
\usepackage{helvet}         
\usepackage{courier}        
\usepackage{type1cm}        
\usepackage{makeidx}         
\usepackage{graphicx}        

\usepackage{multicol}        
\usepackage[bottom]{footmisc}
\usepackage{bbold}

\usepackage{natbib}
\usepackage{amsmath}

\usepackage[
	pdftex=true,
	a4paper,
	pdfpagemode={UseOutlines}, 
	pdfview={Fit},
	pdfstartview={Fit},
	pdftitle={titel},
	pdfpagelabels,
	plainpages=false,
	colorlinks=true,
	linkcolor=darkblu,
	menucolor=darkblue,
	urlcolor=darkblue,
	citecolor=darkblue,
	linktocpage,
	bookmarksnumbered=true,
	bookmarksopen=true,
	hyperindex=true
]{hyperref}

\let\orgautoref\autoref
\renewcommand{\autoref}
        {\def\equationautorefname{Eq.}%
         \def\figureautorefname{Fig.}%
         \def\subfigureautorefname{Fig.}%
         \def\tableautorefname{Tab.}%
         \def\algorithmautorefname{Alg.}%
         \def\appendixautorefname{}%
         \orgautoref}

\makeatletter
\def\tagform@#1{\maketag@@@{\ignorespaces#1\unskip\@@italiccorr}}
\let\orgtheequation\theequation
\def\theequation{(\orgtheequation)}
\makeatother

\begin{document}
	\begin{frontmatter}
		\title{Turbulent jet computations based on MRT and Cascaded Lattice Boltzmann models}
		\author[Brunswick]{S. Geller}\ead{geller@irmb.tu-bs.de}
		\author[Brunswick]{S. Uphoff}\ead{uphoff@irmb.tu-bs.de}
		\author[Brunswick]{M. Krafczyk}\ead{kraft@irmb.tu-bs.de}
		\address[Brunswick]{Institute for Computational Modeling in Civil Engineering, Technische Universit\"at Braunschweig, Germany } 
	
		\begin{abstract}
In this contribution a numerical study of a turbulent jet flow is presented. The simulation results of two different variants of the Lattice Boltzmann method (LBM) are compared. The first is the well-established $D3Q19$ MRT model extended by a Smagorinsky Large Eddy Simulation (LES) model. The second is the $D3Q27$ Factorized Cascaded Lattice Boltzmann (FCLB) model without any additional explicit turbulence model. For this model no studies of turbulent flow with high resolution on nonuniform grids existed so far. The underlying computational procedure uses a time nested refinement technique and a grid with more than a billion DOF. 
 The simulations were conducted with the parallel multi physics solver \textsc{VirtualFluids}. It is shown that both models are feasible for the present flow case, but the FCLB outperforms the traditional approach in some aspects.
    \end{abstract}	
%
		\begin{keyword}
			lattice Boltzmann, large eddy simulation, factorized cascaded Lattice Boltzmann, distributed simulation, jet
		\end{keyword}
\end{frontmatter}

\section{Introduction}
\label{SecIntro}

The Lattice Boltzmann model can be considered an alternative approach to obtain numerical solutions of the Navier-Stokes equations, even though LBM can also be used to investigate finite Knudsen number flows. LBM is based directly on the distribution functions for the particle dynamics of the fluid. The method has successfully been employed to model and simulate a variety of complex fluid flow problems ranging from multi component \cite{Ladd94.1} and multi phase flows  \cite{lee2005stable} to thermal flows \cite{lallemand2003hybrid}, fluid-structure interaction \cite{Geller06.3}, non-Newtonian flows \cite{boyd2006second} and turbulent flows \cite{Teixeira98}.
Over the last years a a number of Lattice Boltzmann variants have been developed to simulate turbulent flows \cite{Teixeira98, Krafczyk03}. 

Even though Direct Numerical Simulation (DNS) is gaining more relevance for certain turbulence flow problems, it is still prohibitively expensive for most relevant applications including turbulence. Any mature CFD scheme should also be capable of incorporating state-of-the-art turbulence models. In the Lattice Boltzmann context large eddy simulation (LES) models are particularly popular due to the small timestep of the explicit scheme and the small overhead needed to implement an algebraic LES model \cite{hou1994lattice,Krafczyk03}, but RANS models have also been used with LBM \cite{Teixeira98}.

 An alternative approach to the simulation of turbulent flows using turbulence models is the use of numerical methods without any explicit turbulence model but relying entirely on a suitable dissipation of the numerical scheme. The fine turbulent scales are not resolved, and the numerical discretization is acting as a filter. Such schemes, named implicit large eddy simulation (ILES) models, are becoming more popular as stated by Grinstein et al. \cite{ILESbuch}. The Factorized Cascaded Lattice Boltzmann (FCLB) model has been shown to give reasonable results at high Reynolds numbers with very low resolution \cite{Geier09} without any explicit turbulence model.  

Our simulations are based on the research code \textsc{VirtualFluids} - a parallel code which is based on MPI and the METIS partitioning tool \cite{Freudiger08}.  A hybrid block data structure to overcome the bottlenecks of the previous approach is used \cite{Freudiger09}. This block data structure enables partitioning of very large datasets because only the block data structure has to be partitioned instead of the entire set of individual nodes. 
For local grid refinement with hierarchical block grids \cite{Freudiger09} the grid refinement strategy of 
Yu et al. \cite{yu2002multi} is employed. See also \cite{yu2003viscous} for a review and evaluation of various refinement techniques and Crouse et al. \cite{crouse2003lb} for applications.

Jet flow is a standard validation problem that has been studied thoroughly both experimentally and numerically, such as in the early experimental work of Wygnanski et al. \cite{wygnanski1968some} and DNS study of Boersma et al. \cite{Boersma1998}. A Lattice Boltzmann study of a turbulent square jet flow has been carried out by Yu et al.\cite{yu2006turbulent,yu2005near}. The MRT (eq. \ref{eqn:MRT}) and SRT (eq \ref{eq:lbgk}) models with Smagorinsky LES have been compared on a uniform grid with a $D3Q19$ stencil, a 19-element stencil in three directions. \cite{menon2004simulation} conducted a further study of a square jet with Lattice Boltzmann LES. 

In this article two different Lattice Boltzmann collision models, namely the $D3Q19$ MRT model with Smagorinsky LES and
the $D3Q27$ Factorized Cascaded Lattice Boltzmann (FCLB) model, are evaluated for their capability to predict turbulent flows for the complex flow case of a free jet. For axisymetrical flows a lack of isotropy has been reported for the $D3Q15$ and $D3Q19$ models, while the $D3Q27$ was found to remove this flaw as White and Chong \cite{white2011rotational} observed when they tested the isotropy of these lattices for flow through a nozzle at $Re\leq 500$ using the BGK and MRT model. They pointed out the importance of reducing isotropy errors as they had found that the errors depended only weakly on the grid resolution. Mayer and H\'azi \cite{mayer2006direct} also observed a lack of isotropy for the $D3Q19$ but not the $D3Q27$ model in a study of laminar and turbulent flow through rod bundles.
 
The article is structured as follows: We start with an overview over different LBM variants, the $D3Q19$ MRT model, the $D3Q27$ models Cascaded Lattice Boltzmann (CLB) and FCLB. The incorporation of large eddy models in LBM is briefly recalled. In the second part of the article we present the testcase of the turbulent jet flow. Firstly, the flow type, for which a semi-analytical solution is known, is described. Next we give a brief description of the experiment to which we compare our data. After that the numerical setup is presented followed by simulation results for the $D3Q19$ MRT model with Smagorinsky LES and the $D3Q27$ FCLB model. 
Finally, the results are discussed and differences between the results from the two approaches are pointed out.

\section{Lattice Boltzmann collision models and subgrid stress model}
\label{LBMbaisicsSGS}
The Lattice Boltzmann scheme emerged in the late 1980's from Lattice Gas Cellular Automata 
\cite{McNamara88} as a new approach to Computational Fluid Mechanics. Unlike conventional
discretizations of the Navier Stokes equations, Lattice Boltzmann equations rely on a discretization
of a simplified Boltzmann equation which is a time-dependent description of the behavior of particle
ensembles. In its simplest form, it is based on a single relaxation time for the non-equilibrium
distribution function \cite{Qian92.1}.
\begin{equation}\label{eq:lbgk}
f_i(x+e_i \Delta t,t+\Delta t)-f_i=\frac{\Delta t}{\tau} ( f_i(x,t)-f_i^{eq}(x,t) )
\end{equation}
for a distribution function $f$,  its equilibrium $f^{eq}$ and the relaxation parameter $\tau$. The components $f_i(x,t)$ of the distribution function depend on the discrete time step $t$, the position $x$ which is related to a discrete node of the numerical grid and the index $i$ for the discrete velocity set. 
The equilibrium for the incompressible model reads \cite{He1997}
\begin{equation}
f_i^{eq}=w_i \left(\delta \rho+3 \frac{  \boldsymbol{\sf e}_i \cdot  \boldsymbol{\sf u}}{c_s^2}+\frac{9}{2} \frac{(  \boldsymbol{ \sf e}_i \cdot  \boldsymbol{\sf u})^2}{c_s^4}-\frac{3}{2} \frac{ \boldsymbol{\sf u}^2}{c_s^2}  \right)
\end{equation}
Here $\delta \rho$ is the density fluctuation for $\rho=\rho_0+\delta \rho$, $u$ is the macroscopic velocity and $c_s$ the speed of sound in the LBM context. By ${\boldsymbol{\sf e}}$ we denote the discretized microscopic velocity.  
For the $D3Q19$ model the weight factors are $w_0=1/3$, $w_1=1/18$, $w_2=1/36$
and for the D3Q27 model $w_0=8/27$, $w_1=2/27$, $w_2=1/54$,  $w_3=1/216$ where $w_3$ is used for the velocity vectors that point to the corners of the cube. The entries of the velocity vectors $ {\boldsymbol{ \sf e}}_i$ for the $D3Q19$ and $D3Q27$ model are: \\

$D3Q19$
	{\scriptsize
	\begin{eqnarray*}
		&&\{\boldsymbol{{\sf e}}_{i}, i=0, \ldots, 18\}=\\
		&&\left\{
		\begin{array}{rrrrrrrrrrrrrrrrrrr}
		0 & c & -c & 0 & 0 & 0 & 0 & c & -c & c & -c & c & -c & c & -c & 0 & 0 & 0 & 0 \\
		0 & 0 & 0 & c & -c & 0 & 0 & c & -c & -c & c & 0 & 0 & 0 & 0 & c & -c & c & -c \\
		0 & 0 & 0 & 0 & 0 & c & -c & 0 & 0 & 0 & 0 & c & -c & -c & c & c & -c & -c & c
		\end{array}
		\right\}
  \end{eqnarray*}
  }

$D3Q27$
	{\scriptsize  
		\begin{eqnarray*}
		&&\{\boldsymbol{{\sf e}}_{i}, i=0, \ldots, 15\}= \\
		&&\left\{
		\begin{array}{rrrrrrrrrrrrrrrr}
		0 & c & -c & 0 & 0 & 0 & 0 & c & -c & c & -c & c & -c & c & -c & 0\\
		0 & 0 & 0 & c & -c & 0 & 0 & c & -c & -c & c & 0 & 0 & 0 & 0 & c \\
		0 & 0 & 0 & 0 & 0 & c & -c & 0 & 0 & 0 & 0 & c & -c & -c & c & c 
		\end{array}
		\right\}
  \end{eqnarray*}
  }
	{\scriptsize  
		\begin{eqnarray*}
		&&\{\boldsymbol{{\sf e}}_{i}, i=16, \ldots, 26\}= \\
		&&\left\{
		\begin{array}{rrrrrrrrrrr}
		 0 & 0 & 0 &c  &-c& c& c& -c&-c& c &-c\\
		 -c & c & -c &c  & c&-c& c& -c& c&-c &-c\\
		 -c & -c & c &c  & c& c&-c&  c&-c&-c &-c
		\end{array}
		\right\}
  \end{eqnarray*}
  }

%

where $c$ is the lattice speed $\Delta t/\Delta x$.
The macroscopic values density $\rho$ and velocity u are computed from summation of the distributions. For the incompressible model we set $\rho=\rho_0+\delta \rho$ and
\begin{eqnarray}
\delta \rho &=& \sum_{i} f_i \\
\boldsymbol{u}_{x} &=& \frac{1}{\rho_0}\sum_{i} f_i \boldsymbol{{\sf e}}_{ix} \\
\boldsymbol{u}_{y} &=& \frac{1}{\rho_0}\sum_{i} f_i \boldsymbol{{\sf e}}_{iy}
\end{eqnarray}
for incompressible models (such as the $D3Q19$ MRT LES model we used in this work) or 
\begin{eqnarray}
\rho &=& \sum_{i} f_i  \label{momZero}\\
\rho \boldsymbol{u}_{x} &=& \sum_{i} f_i \boldsymbol{{\sf e}}_{ix} \\
\rho \boldsymbol{u}_{y} &=& \sum_{i} f_i \boldsymbol{{\sf e}}_{iy} \label{momOne}
\end{eqnarray}
for compressible models (such as the $D3Q27$ FCLB model used in this work).
More advanced approaches use Multiple Relaxation Times (MRT) \cite{Humieres02} where the relaxation step takes place in moment space. 
To introduce moments let us first define an expectation value of a linear operator B acting on distribution functions f in the discretized velocity space
\begin{equation}
\left\langle B \right \rangle=\sum_i \left( B(f)\right)_i
\end{equation}
Moments are then defined as expectation values of powers of the discrete velocities
\begin{equation}
\mu_{x^i y^j z^k}=\left\langle e_x^i e_y^j e_z^k \right \rangle
\end{equation}
in accordance with the definitions for distribution functions in continuous spaces that can be found in e.g. \cite{severini2005elements}.
An alternative notation that avoids multiple subscripts is 
\begin{equation}
\mu_{\underbrace{1\ldots1 }_i\,\underbrace{ 2\ldots2}_j \, \underbrace{3\ldots3}_k}:=\mu_{x^i y^j z^k}
\end{equation}
The density and momentum defined in eqns. \ref{momZero} to \ref{momOne} are the moments of order zero and one. The transformation from distribution functions to moments is a linear transformation M. The Lattice Boltzmann relaxation step for the MRT model is then given by the equation
\begin{equation}\label{eqn:MRT}
f_i(x+e_i,t+\Delta t)=f_i+\mathbf{M}^{-1} \mathbf{S} (\mathbf{M} f(x,t)-m^{eq}(x,t))
\end{equation}
with $m^{eq}=M f^{eq}$. The collision parameters $s_{ii}$ of the diagonal matrix $S$ are chosen such that the density and momentum are conserved and that the viscosity is correctly represented using
\begin{equation}
	\label{eq:relaxationTau}
	\tau=3\frac{\nu }{c_s^{2}}+\frac{1}{2}\Delta t,
\end{equation}%
where $\nu $ is the kinematic viscosity in lattice units $\left[ \nu \right]=\Delta x^2/ \Delta t$.
Additional free parameters can be chosen as to improve the stability of the model. 
The moments, or central moments, belong to different invariant subsets of the stencils' symmetry groups as described in reference \cite{rubinstein2008theory}. They have to be relaxed with one relaxation factor each (definition see below). Table  1
lists these groups and the corresponding relaxation factors. For the $D3Q19$ model only the moments for $s_1$, $s_2$, $s_3$, $s_4$, and $s_7$ are considered. 
We chose the values of the relaxation parameters to be $s_i=1 \,  \forall i \neq 1$ and $s_1=\Delta t/ \tau$ for both the $D3Q19$ and the $D3Q27$ simulation runs.
 
\begin{table}[h]
\label{tab:groupsOfRelParams}
\centering
\caption{relaxation factors}
\begin{tabular}{| l | c |}
  \hline
moment & relaxation parameter\\
  \hline
$\mu_{12}$, $\mu_{13}$, $\mu_{23}$, $\mu_{11}-\mu_{22}$, $\mu_{11}-\mu_{33}$&  $s_1$\\
$\mu_{11}+\mu_{22}+\mu_{33}$ &  $s_2$\\
$\mu_{122}+\mu_{133}$, $\mu_{112}+\mu_{233}$, $\mu_{133}+\mu_{233}$ & $s_3$\\
$\mu_{122}-\mu_{133}$, $\mu_{112}-\mu_{233}$, $\mu_{133}-\mu_{233}$ & $s_4$\\
$\mu_{123}$ & $s_5$ \\ 
$\mu_{1122}-2\mu_{1133}+\mu_{2233}$, $\mu_{1122}+\mu_{1133}-2\mu_{2233}$ & $s_7$\\ 
$\mu_{1122}+\mu_{1133}+\mu_{2233}$ & $s_7$\\ 
$\mu_{1223}$, $\mu_{1223}$, $\mu_{1233}$ & $s_8$ \\
$\mu_{12233}$, $\mu_{11233}$, $\mu_{11223}$ & $s_9$ \\
$\mu_{112233}$ & $s_{10}$ \\
\hline
\end{tabular}
\end{table}
The first order moments and the density do not appear in table 1 
as they are conserved.
\newline A further development are the cascaded Lattice Boltzmann schemes. The so-called Cascaded Lattice Boltzmann (CLB) method was developed by Geier et al. \cite{Geier06}. Further developments have been made
 to introduce different equilibria \cite{Geier09} which constitute the FCLB scheme. All CLB-methods rely on the basic idea to use central moments instead of uncentered moments and to use lower order moments after relaxation for the computation of the higher order moments (hence the term cascaded). Central moments are defined as
\begin{equation}
M_{x^i}^c=\left\langle \left( x-\left\langle x\right\rangle \right)^i   \right\rangle
\end{equation}
for the expectation value 
$\left\langle x \right\rangle$ of a function $f(x)$. In our case, the expectation value is intended 
for the discrete distribution function 
$f(\vec{x},\vec{\mu},t)$ with respect to momentum space as defined above. 
For three directions we have a product of one-dimensional terms.
\begin{equation}
M^c_{{e_{x}}^i {e_{y}}^j {e_{z}}^k}=\left\langle \left( e_{x} -\left\langle e_{x} \right\rangle \right)^i   \left(e_{y} -\left\langle e_{y} \right\rangle \right)^j \left(e_{z} -\left\langle e_{z} \right\rangle \right)^k\right\rangle
\end{equation}
Intuition leads us to suspect that non-linear operations should be carried out in the inertial frame. Consider for example the second order central moments (i.e. the variances). 
The uncentered moment is $\mu_{11}=v_x^2+var(v_x)$. Hence the term $\sum_i f^{eq}_i e_{ix} e_{ix}-\rho/3=v_x^2$ has been removed by the transformation and the central moment then is the variance only. 
The equilibrium central moments are chosen as the corresponding central moments of the Gauss function where the variance is the speed of sound $c_s$. These are the same equilibria as those obtained from taking the central moments of the MRT-equilibria if third-order terms are taken into account for the MRT equilibria as well. The Factorized CLB method is a special CLB method which aims at removing the influence of the lower-order central moments on the fourth- and higher order moments at an acceptable computational cost. This correction leads to an improved stability of the method and further reduces errors with respect to isotropy that occur with any finite stencil \cite{Geier09}. The transformation and specific equilibria for the $D3Q27$ stencil are given in table \ref{tab:fclb}.
The original implementation of  Geier et al. \cite{Geier06, Geier09} computed the changes in the moments after collision. Our implementation differs from the original implementation as we do not compute the change in the moments, but recompute the entire moments. The basis for the moments used in reference \cite{Geier06} has some differences from the basis used here.



\begin{table}
\label{tab:fclb}
\centering
\caption{transformation to central moments}
\begin{tabular}{  | l | c  |r  |}
  \hline
central moment & transformation & equilibrium \\
  \hline
$M_{11}^c$ & $\mu_{11}-\mu_1^2$ & $1/3$  \\  
\hline
$M_{22}^c$ & $\mu_{22}-\mu_2^2$ & $1/3$ \\  
\hline
$M_{33}^c$ & $\mu_{33}-\mu_3^2$ & $1/3$ \\  
\hline
$M_{12}^c$ & $\mu_{12}-\mu_1 \mu_2$ & $0$ \\  
\hline
$M_{13}^c$ & $\mu_{13}-\mu_1 \mu_3$ & $0$ \\  
\hline
$M_{23}^c$ & $\mu_{23}-\mu_2 \mu_3$ & $0$ \\  
\hline
$M_{112}^c$ & $\mu_{112}-\mu_{11} \mu_2 -2\mu_1 \mu_{12}+2\mu_1 \mu_{1} \mu_2$ & $0$ \\ 
\hline
$M_{122}^c$ & $\mu_{122}-\mu_{22} \mu_1 -2\mu_2 \mu_{12}+2\mu_1 \mu_{2} \mu_2$ & $0$ \\ 
\hline
$M_{113}^c$ & $\mu_{113}-\mu_{11} \mu_3 -2\mu_1 \mu_{13}+2\mu_1 \mu_{1} \mu_3$ & $0$ \\ 
\hline
$M_{133}^c$ & $\mu_{133}-\mu_{33} \mu_1 -2\mu_3 \mu_{13}+2\mu_1 \mu_{3} \mu_3$ & $0$ \\ 
\hline
$M_{223}^c$ & $\mu_{223}-\mu_{22} \mu_3 -2\mu_2 \mu_{23}+2\mu_2 \mu_{2} \mu_3$ & $0$ \\ 
\hline
$M_{233}^c$ & $\mu_{233}-\mu_{33} \mu_2 -2\mu_3 \mu_{23}+2\mu_2 \mu_{3} \mu_3$ & $0$ \\ 
\hline
$M_{123}^c$ & $\mu_{123}-\mu_{12} \mu_{3}-\mu_{23} \mu_{1}-\mu_{3} \mu_{12}+2\mu_{1} \mu_{2} \mu_{3}$ & $0$ \\ 
\hline
$M_{1122}^c$ & $\mu_{1122}-2\mu_{112} \mu_2-2\mu_{122} \mu_1 +4 \mu_{11} \mu_{22}$& \\ 
& $+\mu_{1}^2  \mu_{22}+\mu_{11} \mu_{2}^2+4\mu_{1}\mu_2 \mu_{12}-3\mu_{1}^2 \mu_{2}^2$ &   $M_{11}^c M_{22}^c$\\ 
\hline
$M_{1133}^c$ & $\mu_{1133}-2\mu_{113} \mu_3-2\mu_{133} \mu_1 +4 \mu_{11} \mu_{33}$&\\ 
 &$+\mu_{1}^2  \mu_{33}+\mu_{11} \mu_{3}^2+4\mu_{1}\mu_3 \mu_{13}-3\mu_{1}^2 \mu_{3}^2$ &    $M_{11}^c M_{33}^c$\\ 
\hline
$M_{2233}^c$ & $\mu_{2233}-2\mu_{223} \mu_3-2\mu_{233} \mu_2 +4 \mu_{22} \mu_{33}$& \\ 
 &$+\mu_{2}^2  \mu_{33}+\mu_{22} \mu_{3}^2+4\mu_{2}\mu_3 \mu_{23}-3\mu_{2}^2 \mu_{3}^2$ &   $M_{22}^c M_{33}^c$\\ 
\hline
$ M_{1233}^c$& $ -3 \mu_3^2 \mu_2 \mu_1+\mu_{33} \mu_2 \mu_1 + 
            2 \mu_3 \mu_{23} \mu_1 -     \mu_{233} \mu_1 +$ &\\
            & 
            $2 \mu_3\mu_2 \mu_{13} -     \mu_2 \mu_{133} + 
            \mu_3^2 \mu_{12} - 2 \mu_3 \mu_{123} + \mu_{1233}$ &$M_{33}M_{12}$\\
 \hline          
$ M_{1223}^c$& $ -3 \mu_2^2 \mu_3 \mu_1+\mu_{22} \mu_3 \mu_1 +  
            2 \mu_2 \mu_{23} \mu_1 -     \mu_{223} \mu_1 + $ &\\
            & 
            $2 \mu_3\mu_2 \mu_{12} -     \mu_3 \mu_{122} + 
            \mu_2^2 \mu_{13} - 2 \mu_2 \mu_{123} + \mu_{1223}$ &$M_{22}M_{13}$\\
\hline
$ M_{1123}^c$& $ -3 \mu_1^2 \mu_2 \mu_3+\mu_{11} \mu_2 \mu_3 +  
            2 \mu_3 \mu_{12} \mu_1 -     \mu_{112} \mu_3 +$ &\\
            & 
            $2 \mu_1\mu_2 \mu_{13} -     \mu_2 \mu_{113} + 
            \mu_1^2 \mu_{32} - 2 \mu_1 \mu_{123} + \mu_{1123}$ &$M_{23}M_{11}$\\
 \hline           
$M_{11223}^c$ & $4 \mu_3 \mu_2^2 \mu_1^2 - 2 \mu_2 \mu_{23} \mu_1^2 -\mu_3 \mu_{22} \mu_1^2 $ &\\
  & 
  $ + \mu_{223} \mu_1^2 - 
  2\mu_2^2 \mu_1 \mu_{13} - 
  4\mu_3 \mu_2 \mu_1 \mu_{12} + 
  4\mu_2 \mu_1 \mu_{123} +$   &\\
  & 
  $2\mu_3 \mu_1 \mu_{122} - 2 \mu_{1} \mu_{1223} - 
  \mu_3 \mu_2^2 \mu_{11} + \mu_2^2 \mu_{113} + $ &\\
  & 
  $2\mu_3 \mu_2 \mu_{112} - 2 \mu_2 \mu_{1123} - 
  \mu_3 \mu_{1122} + \mu_{11223}$  &$0$ \\
\hline
$M_{11233}^c$ & $4 \mu_3^2 \mu_2 \mu_1^2 - 
  \mu_{33} \mu_2 \mu_1^2 - 2 \mu_3 \mu_{23} \mu_1^2$ &\\
  & 
  $+  \mu_{233}\mu_1^2 - 
  4 \mu_3 \mu_2 \mu_1 \mu_{13} + $ & \\
  & 
  $2 \mu_2 \mu_1 \mu_{133} - 
  2 \mu_3^2 \mu_1 \mu_{12} + $ &\\
  & 
  $4 \mu_3 \mu_1 \mu_{123}-2 \mu_1 \mu_{1233} - 
  \mu_3^2 \mu_2 \mu_{11} + $ & \\
  &
  $2 \mu_3 \mu_2 \mu_{113} - \mu_2 \mu_{1133} + 
  \mu_3^2 \mu_{112} - 2 \mu_3 mu_{1123} +\mu_{11233} $ &$0$ \\
\hline  
$M_{12233}^c$ & $4 \mu_3^2 \mu_2^2 \mu_1 - 
  \mu_{33} \mu_2^2 \mu_1 - 2 \mu_3 \mu_{13} \mu_2^2 $ &\\
  & 
  $+  \mu_{133}\mu_2^2 - 
  4 \mu_3 \mu_2 \mu_1 \mu_{23} + 
  2 \mu_2 \mu_1 \mu_{233} - $ &\\
  &
  $  2 \mu_3^2 \mu_2 \mu_{12} + 
  4 \mu_3 \mu_2 \mu_{123}-2 \mu_2 \mu_{1233} -$ &\\
  & 
  $\mu_3^2 \mu_1 \mu_{22} + 
  2 \mu_3 \mu_1 \mu_{223} - \mu_1 \mu_{2233} + 
  \mu_3^2 \mu_{122} - 2 \mu_3 mu_{1223} +\mu_{12233} $ &$0$ \\

\hline
 $M_{112233}^c$&$
            -5 \mu_3^2 \mu_2^2 \mu_1^2 + 
            \mu_{33}\mu_2^2 \mu_1^2 + 4 \mu_3 \mu_2 \mu_{23} \mu_1^2 - 
            2 \mu_2 \mu_{233} \mu_1^2 +$ &\\
            & 
            $\mu_3^2 \mu_{22} \mu_1^2 - 
            2 \mu_3 \mu_{223} \mu_1^2 + 
            \mu_{2233} \mu_1^2+ 
            4 \mu_3 \mu_2^2 \mu_1 \mu_{13} -$ &\\
            & 
            $2 \mu_2^2 \mu_1 \mu_{133} + 
            4 \mu_3^2 \mu_2 \mu_1 \mu_{12} - 
            8 \mu_3 \mu_2 \mu_1 \mu_{123} + 
            4 \mu_2 \mu_1 \mu_{1233} -  $ &\\
            & 
            $2 \mu_3^2 \mu_1 \mu_{122} +4 \mu_3 \mu_1 \mu_{1223}-2 \mu_1 \mu_{12233} + 
            \mu_3^2 \mu_2^2 \mu_{11} $  &\\
            & 
            $ - 
            2 \mu_3\mu_2^2 \mu_{113} + 
            \mu_2^2 \mu_{1133} -2 \mu_3 \mu_3 \mu_2 \mu_{112} + 
            4 \mu_3 \mu_2 \mu_{1123}$ & \\
            &
             $-2 \mu_2 \mu_{11233} + 
            \mu_3^2 \mu_{1122} - 2 \mu_3 \mu_{11223} + 
            \mu_{112233} $ & $M_{11}^cM_{22}^cM_{33}^c$ \\
 \hline
\end{tabular}
\end{table}
We chose this implementation because of its more modular properties. The first transformation is the same as for the MRT model. The less compressed implementation is less prone to errors and makes it easier to change algorithmic details later. On the other hand, it is not as optimized as the original version with respect to the number of floating point operations (FLOPS). A large number of FLOPS can be eliminated if relaxation parameters are fixed.
The CLB and FCLB model are suspected to have ILES capabilities. This has been subject to investigation in references \cite{Geier06}, \cite{Geier09} and \cite{Geier08.1} where no additional turbulence model was used. For under-resolved simulations of turbulent flows with the LBGK or MRT model, however, a turbulence model is needed. 
\newline The standard Smagorinsky model is a popular choice due to its simplicity and efficiency. In this model the eddy viscosity $\nu_{\tau}$ depends only on the magnitude of the strain rate ${\bf {S}}$ and the grid spacing $\Delta x$
\begin{equation}
	\label{eqSmago}
		\nu_{\mathrm{T}}=(C_S\Delta x)^2\|{\mathbf{S}}\|
\end{equation} 
where the strain rate tensor is defined as
\begin{equation}
	 S_{\alpha \beta} =\frac{1}{2} \left(\frac{\partial \bar{u}_\alpha}{\partial x_\beta}+\frac{\partial \bar{u}_\beta}{\partial x_\alpha}\right).
\end{equation}
and the Smagorinsky constant $C_S$. We chose $C_S=0.18$, which is in the range of values suggested by Rogallo and Moin \cite{rogallo1984numerical}. In the Lattice Boltzmann context the viscosity $\nu$ is related to the relaxation time $\tau$ as defined in equation \ref{eq:relaxationTau}. The norm of the strain rate can be computed locally from
\begin{equation}
	\label{eqStrainRate}
	\left\| {\mathbf{S}} \right\|=-\frac{3}{2\tau_{\mathrm{total}}c^2} 	\left\|\Pi^{\mathrm{neq}}  \right\|,
\end{equation}
where the norm of the momentum flux tensor $\Pi^{\mathrm{neq}}$ is defined as 
\begin{equation}
\left\|	\Pi^{\mathrm{neq}} _{\alpha\beta} \right\|=
\left(\sum_{\alpha,\beta} \left(\mu_{\alpha \beta}^{neq}-\delta \rho/3 \delta_{\alpha \beta} \right)^2 \right)^{1/2}
\end{equation}
in the case of an incompressible model.
The total relaxation factor can be obtained from the following equation \cite{hou1994lattice} 
\begin{equation}
\label{eqTauTotFinal}
	\tau_{\mathrm{total}}=\frac{3}{c^2}\nu_{\mathrm{0}}+\frac{1}{2}\Delta t+\frac{\sqrt{\tau_0^2+\frac{18C_S^2{\Delta t}^2 Q}{c^2}}-\tau_0}{2}
\end{equation}
where 
\begin{equation}
\label{EqnQ}
Q=\sqrt{\sum_{\alpha\beta}2\Pi^{\mathrm{neq}}_{\alpha\beta}\Pi^{\mathrm{neq}}_{\alpha\beta}}.
\end{equation}
Note that the procedure is entirely local. No information from adjacent nodes is required, which is highly desirable for parallel computations.
For the description of the hierarchical block structured grid approach for the SGS model
we refer to reference \cite{stiebler20113475}.

\section{Validation of turbulent jet flow}
\label{SecTestCase}

A turbulent jet at $Re=6760$ based on the size of the orifice and the inflow velocity is simulated using the FCLB method with the $D3Q27$ stencil and with the MRT model with Smagorinsky LES and the $D3Q19$ stencil. The simulation results are compared to experimental data from Ming et al. \cite{Ming01}. The section is structured as follows: Firstly, the experimental setup is described. The setup of the numerical solution is described after that, followed by the results of the simulations. Finally, the results are discussed and differences between the results from the two approaches are pointed out.

\subsection{Experimental setup}
The simulations are based on an experiment described in reference \cite{Ming01}. The properties of a turbulent jet at a Reynolds number of $6760$ based on the size of the opening of $4mm$ and on the inflow velocity of $1.69 m/s$ were measured using Doppler laser anemometry. The experiment was carried out in a water tank of $6m$ length in flow direction, $0.2m$ width and $0.4m$ height. The tank is open and the jet enters the tank through a nozzle. At the back of the water tank a drain is present to keep the water level constant.


\subsection{Numerical setup}
With the numerical setup we try to mimic the experimental setup as closely as possible. We use the same size of domain in horizontal, vertical, and spanwise direction. Solid boundaries are modeled by no-slip boundaries. The air-water interface at the upper boundary is modeled by a free-slip condition because a free-surface condition would pose a major additional computational effort and the effect of the wave generation is considered to be negligible for this testcase. Instead of the weir outflow we set a fixed pressure boundary condition. The nozzle was positioned at $0.5 m$, approximated as a cylinder with second-order accurate interpolated no-slip walls \cite{Bouzidi01}. 
The point of origin is on the bottom, left, frontal corner of the basin.
Instead of the nozzle used in the experiment a cylinder is inserted, which extends from (0.0, 0.1, 0.2) to (0.5, 0.1, 0.2) meters and has a radius of $2mm$. On the right emitting end of the cylinder a constant inflow velocity is defined.
The boundary condition at the walls is a noslip condition.

For the discretization of the domain a hybrid block structured grid with a hierarchical refinement structure is used.
Due to the geometrical refinement a nested time step approach is used leading to a globally constant CFL number for the distributions.
The refinement and coarsening strategy is described in \cite{Geller06.3,Freudiger09,Geller10}.
Seven levels of refinement are used to discretize this setup.
The grid resolution is $0.0947 mm$ on the finest and $6.06051 mm$ on the coarsest level.
So the nozzle with its $4 mm$ of diameter is discretized with $42.24$ nodes in the finest domain.
The timestep varies between $0.000103522 s$ (coarse) and $0.0000016 s$ (fine).
The domain is resolved with $83522$ blocks, each of which corresponds to nodal matrix of the size 11x11x11.
In sum $111$ million grid nodes are used. This means three billion degrees of freedom for the $D3Q27$ model and 2.1 billion for the $D3Q19$ model.
The domain was decomposed for parallelization with the METIS library \cite{Karypis98}.

\begin{figure}[htp]
	\centering
		\includegraphics[width=0.80\textwidth]{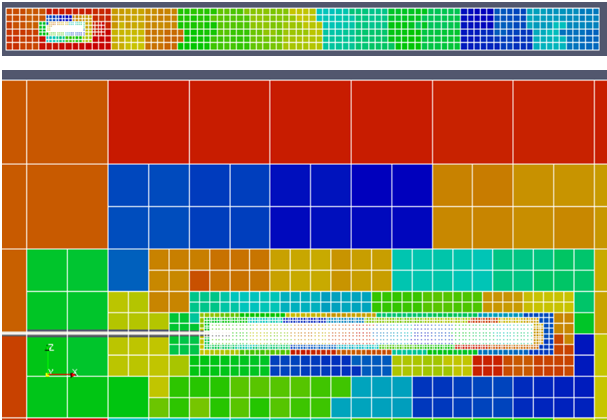}
	\caption{partial view of the domain discretization with blocks of 11x11x11 nodes each , the color indicates the subdomain index after decomposition with METIS}
	\label{fig:discretization}
\end{figure}

\begin{figure}[htp]
	\centering
		\includegraphics[width=0.80\textwidth]{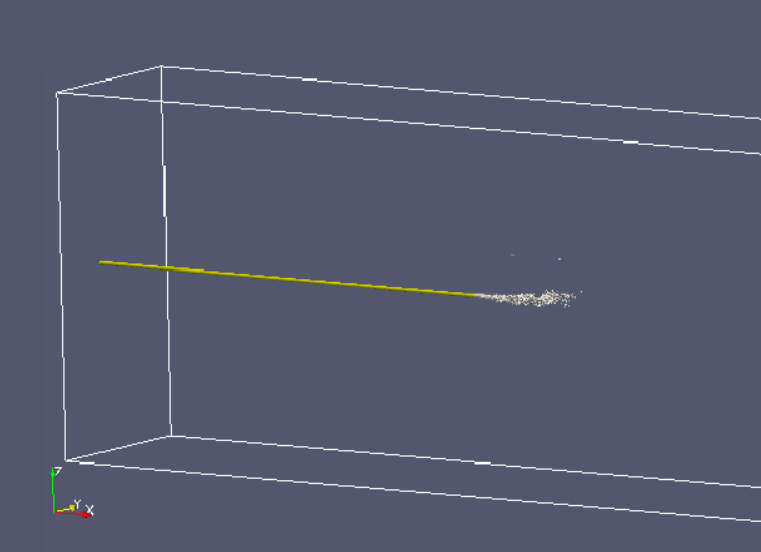}
	\caption{system: pipe with velocity contour at $0.5m/s$ after time $t=3.9s$}
	\label{fig:system}
\end{figure}

The physical parameters are the fluid density	of $998.2 \;  kg/m^3$,
the kinematic viscosity of $10^{-6} \; m^2/s$, the Reynolds number (Re) $6760$ 
(related to the nozzle diameter and inflow speed), and the computation time which covered $3.9 s$ real time.


\subsection{Results}
\begin{figure}[htp]
	\centering
		\includegraphics[width=0.80\textwidth]{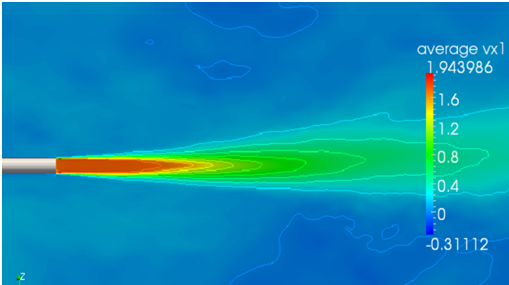}
	\caption{averaged horizontal velocity behind the nozzle, D3Q19 MRT LES }
	\label{fig:aveVeloD3Q19}
\end{figure}

\begin{figure}[htp]
	\centering
		\includegraphics[width=0.75\textwidth]{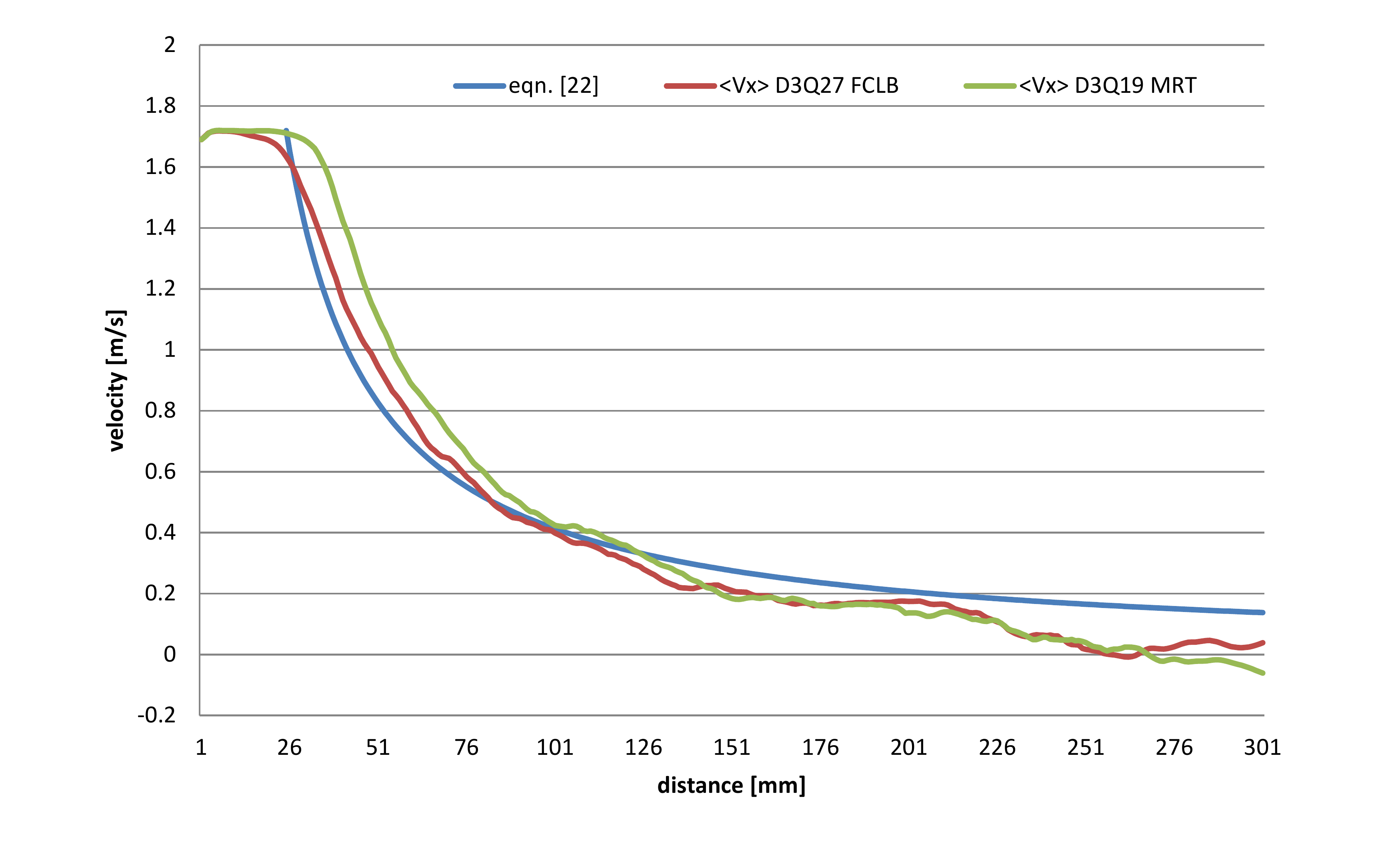}
	\caption{averaged velocity along the jet axis }
	\label{fig:aveVeloJetAxis}
\end{figure}

\begin{figure}[htp]
	\centering
		\includegraphics[width=0.95\textwidth]{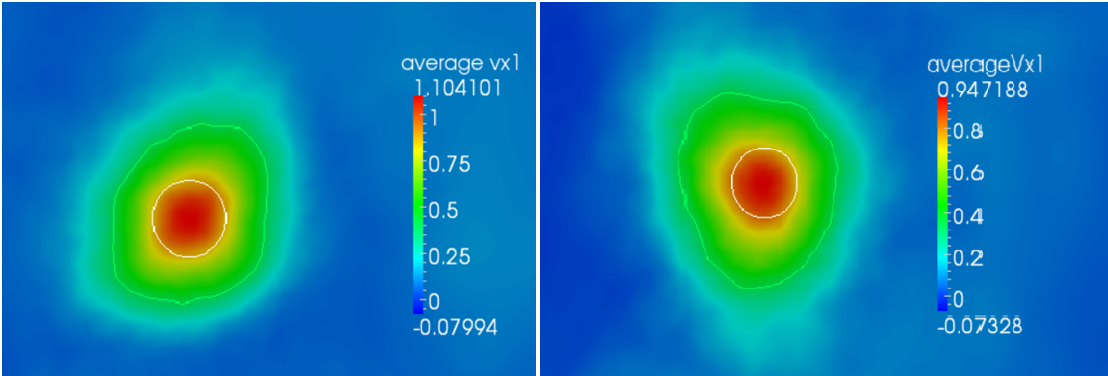}
	\caption{averaged velocity orthogonal to jet axis $5 cm$ behind nozzle, left D3Q19 MRT LES, right D3Q27 FCLB, 
	contour at $0.368 u_m$ }
	\label{fig:aveVelo5cm}
\end{figure}

We compare the averaged velocity along the axis abtained for the FCLB and for the MRT model with the semi-analytical results from \cite{Ming01}. Figure \ref{fig:aveVeloJetAxis} shows a good match for both models. 
Pictures \ref{fig:aveVeloD3Q19} and \ref{fig:aveVelo5cm} give a qualitative idea of the flow dynamics. Immediately behind the opening the flow field is laminar. As eddies develop in the shear layer between jet and surrounding flow, the jet becomes wider with increasing distance from the nozzle.


\begin{figure}[htp]
	\centering
		\includegraphics[width=0.80\textwidth]{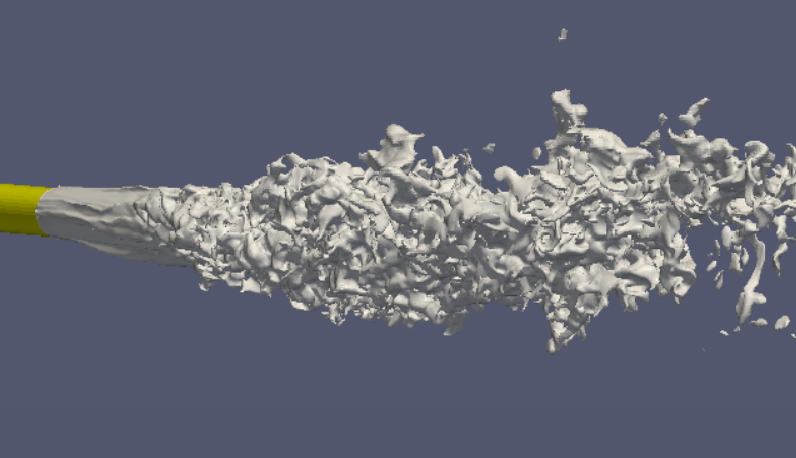}
	\caption{contour of the velocity at $0.5m/s$ after time $t=3.9s$}
	\label{fig:veloContour05ms}
\end{figure}

According to  Ming et al. \cite{Ming01} the average axial velocity behind the nozzle can be described as:

\begin{equation}
	\label{eq:aveJetSpeed}
	 u_m = u_0 \, k_{\parallel} \, \frac{D}{x}
\end{equation}

with nozzle diameter $D = 4 mm$, the distance $x$ from the nozzle,
the averaged velocity $u_m$ at position $x$,
and the inflow speed $u_0=1.69 m/s$. The constant $k_{\parallel}$ has to be determined experimentally and was determined to $k_{\parallel}=6.104$ for the present setup.

The spreading width $b_g$ for the jet is defined as the half-width of the velocity over the distance from the jet axis at a given distance from the jet assuming a Gaussian shape. This leads to a velocity $u_x = 0.368 u_m$ which is present at half the spread-width from the jet axis. The spreading width grows linearly with the distance from the jet \cite{Ming01}
\begin{equation}
	\label{eq:spreadFunction}
	 b_g = k_{\perp} x
\end{equation}
 Ming et al. \cite{Ming01} found a value of $k_{\perp}=0.109$.

\begin{figure}[htp]
	\centering
		\includegraphics[width=0.90\textwidth]{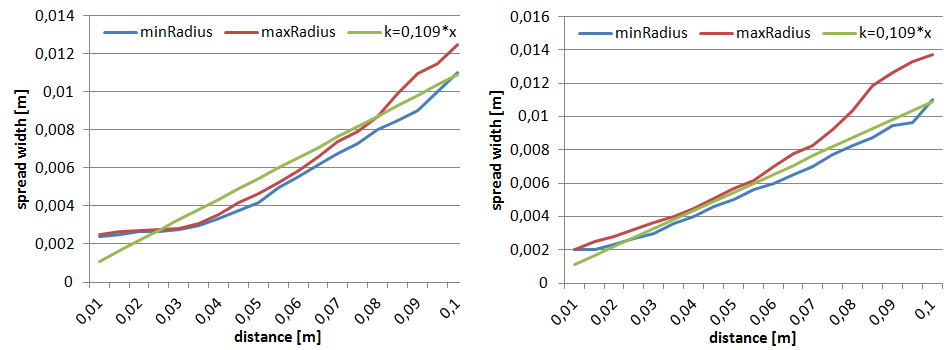}
	\caption{spreading width, left D3Q19 MRT LES, right D3Q27 FCLB }
	\label{fig:spreadwidth}
\end{figure}

Due to the asymmetric behavior of the jet, the minimum and maximum radius
of the spreading function for the distance to the isoline of constant velocity $u_x$ at $u_x = 0.368 u_m$
is given in figure \ref{fig:spreadwidth}.

\begin{figure}[htp]
	\centering
		\includegraphics[width=0.95\textwidth]{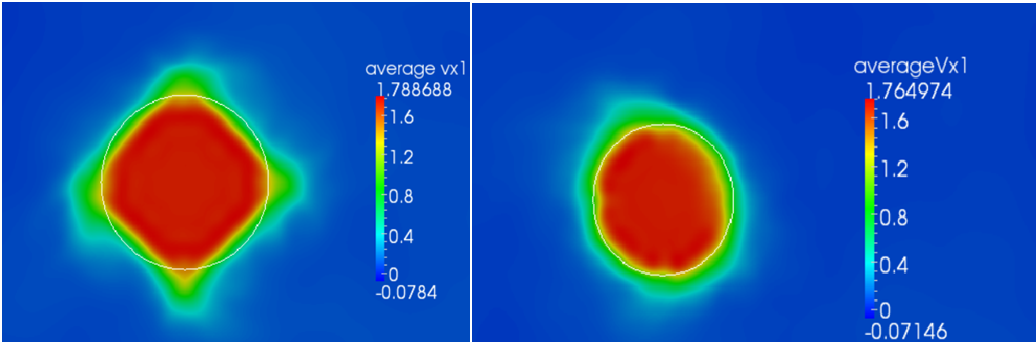}
	\caption{averaged velocity profile $1.5 cm$ behind nozzle, left D3Q19 MRT LES, right D3Q27 FCLB }
	\label{fig:aveVeloProfile1.5cm}
\end{figure}

\begin{figure}[htp]
	\centering
		\includegraphics[width=0.95\textwidth]{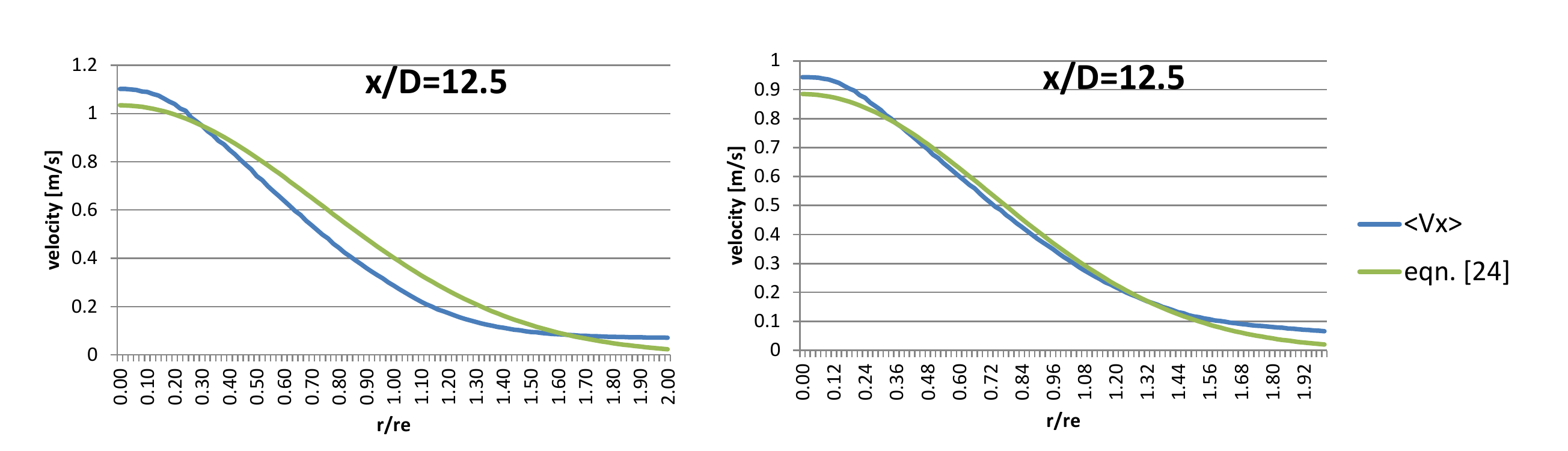}
	\caption{averaged velocity profile $5 cm$ behind nozzle, left D3Q19 MRT LES, right D3Q27 FCLB }
	\label{fig:aveVeloProfile5cm}
\end{figure}

\begin{figure}[htp]
	\centering
		\includegraphics[width=0.95\textwidth]{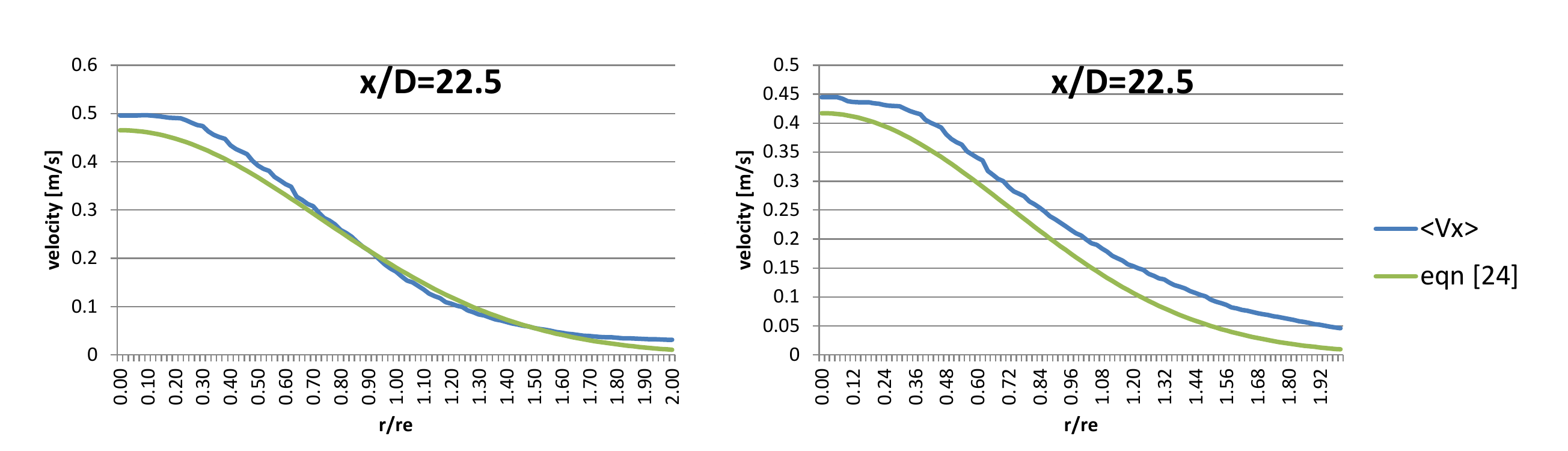}
	\caption{averaged velocity profile $9 cm$ behind nozzle, left D3Q19 MRT LES, right D3Q27 FCLB }
	\label{fig:aveVeloProfile9cm}
\end{figure}

\begin{figure}[htp]
	\centering
		\includegraphics[width=0.95\textwidth]{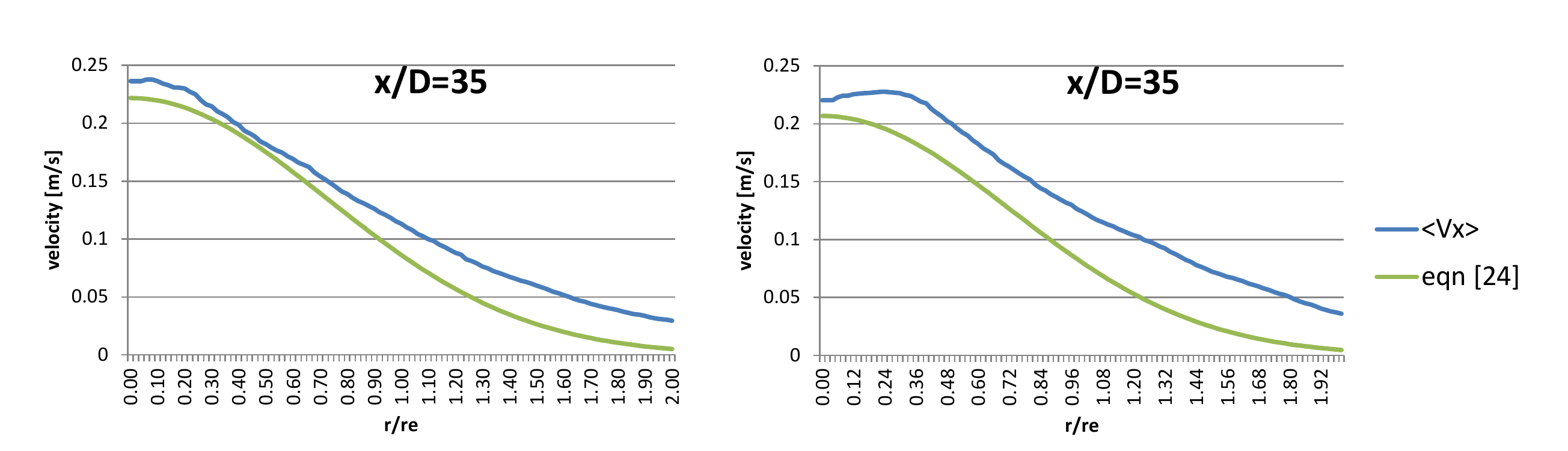}
	\caption{averaged velocity profile $14 cm$ behind nozzle, left D3Q19 MRT LES, right D3Q27 FCLB }
	\label{fig:aveVeloProfile14cm}
\end{figure}

\begin{figure}[htp]
	\centering
		\includegraphics[width=0.95\textwidth]{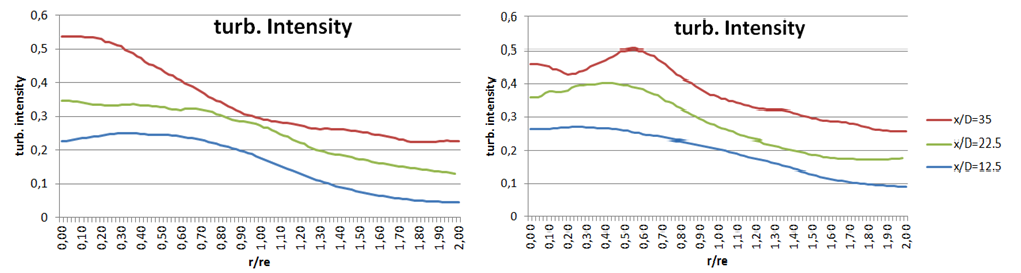}
	\caption{turbulent intensity orthogonal to jet axis, left D3Q19  MRT LES, right D3Q27 FCLB }
	\label{fig:turbIntensity}
\end{figure}

The velocity distribution orthogonal to the jet axis is determined according to \cite{Ming01} by:

\begin{equation}
	\label{eq:veloDistribution}
	 u_x = u_m \cdot 0.938 \cdot e^{-0.944 \left( r/r_e \right)^2}
\end{equation}

with the averaged velocity at jet axis $u_m$, the velocity $u_x$ at the position $r$,
the radius $r_e$ where $u_x = 0.368 u_m$, and radius $r$. The constants have again been determined experimentally.
For different distances behind the nozzle figures \ref{fig:aveVeloProfile5cm},
\ref{fig:aveVeloProfile9cm} and \ref{fig:aveVeloProfile14cm} show the computed results
in comparison with the semi-analytical solution. For the computation of the averaged velocities as well as the turbulent intensity, several lines in different directions from the jet center orthogonal to the jet axis are averaged in addition to averaging in time.
In figure \ref{fig:turbIntensity} the distribution of the turbulent intensity at different positions is shown which was determined from

\begin{equation}
	\label{eq:turbIntensity}
	 tI = \frac{\sqrt{ \left\langle\left( u_{x}  - \left\langle u_x \right \rangle\right)^2 \right\rangle}}{u_{m,x}}
\end{equation}

\begin{figure}[htp]
	\centering
		\includegraphics[width=0.5\textwidth]{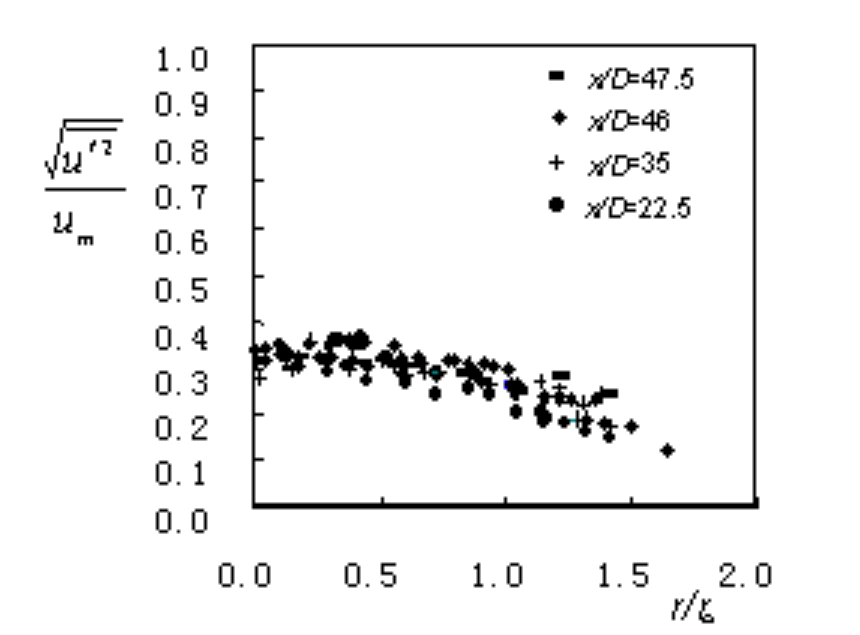}
	\caption{experimental data for the turbulent intensity, from \cite{Ming01}}
	\label{fig:turbIntensityExp}
\end{figure}

Figures \ref{fig:aveVeloJetAxis} to \ref{fig:aveVeloProfile14cm} show the computed results for the two models in comparison with the semi-analytical solution. 
As can be seen from figure \ref{fig:aveVeloJetAxis}, the $D3Q27$ FCLB model is slightly more successful at reproducing the velocity profile along the jet centerline than the $D3Q19$ MRT model with Smagorinsky LES.  The same is true for the spreading width (fig. \ref{fig:spreadwidth}) for moderately large distances from the nozzle. 
For distances larger than $7 cm$ the error in the spreading width of the FCLB model grows, but this may be due to the limited averaging time of 0.9 seconds. The same behavior is observed for the average velocity profiles normal to the jet axis. We believe that the excessive eddy viscosity that occurs with the Smagorinsky LES model in shear layers delays the transition to turbulence.


 An interesting observation is that the mean velocity contours normal to the jet axis diverge from the expected circular shape for the D3Q19 model. The discretization of the velocity space with $19$ vectors seems insufficient to reproduce this particular flow feature. The use of the $D3Q27$ FCLB model improves the isotropy of the flow field. Similar effects have been observed previously by White and Chong \cite{white2011rotational} in a comparison of $D3Q19$ and $D3Q27$ BGK-type models at Reynolds numbers up to $Re=500$. \cite{Geier09} shows a comparison between different stencils and collision models for laminar flows and  also found that the $D3Q27$ FCLB model showed the least anisotropy among the models studied.
\section{Conclusion}
\label{Conclusions} 

In this paper we presented a comparison of a $D3Q19$ MRT model with Smagorinsky LES and the $D3Q27$ FCLB model. We demonstrated that both models correctly reproduce the dynamics of turbulent jet flow. 
The computation of one second real time on 395 cores took two days.
The decay of the axial velocity is in good agreement with the semi-analytical solution.
The solution from D3Q27 FCLB model matches the semi-analytical result better than the D3Q19 LES model.
In the range of $0 \, cm$ to $10 \, cm$ behind the nozzle the spreading functions are in good agreement with the empirical relation determined from experiments.
The velocity profile of a cross-section matches the Gauss function obtain from empirical relations well.
One important aspect is that the D3Q19 LES model shows notable anisotropies whereas
the D3Q27 FCLB model shows no such defect.

We conclude that the Lattice Boltzmann method is suitable for jet induced turbulent incompressible flows even with a simple turbulence model (LES) and an enhanced model (FCLB) used in this work.
The potential of the FCLB model for computing turbulent flows is demonstrated.  
\section{Acknowledgements}
\label{Acknowledgements} 

The authors appreciate the support of the `Federal Waterways 
Engineering and Research Institute' and valuable discussions with Prof.\ S\"ohngen and Mr.\ Spitzer.
In addition, Sonja Uphoff acknowledges financial support from German research foundation (DFG) from the collaborative research project SFB 880.

\bibliographystyle{elsarticle-num}
\bibliography{cm_full,mybib}

\end{document}